| Summary | Self-organized patterns (SOP) in plasma discharges arise from the complex interplay of electric field, reactive species and charged particles, driven by non-linear plasma dynamics. While studies have explored SOP formation in various configurations, no systematic comparison of positive and negative DC glow discharges has been conducted to explain why SOP form exclusively when polarization is negative. This study aims to analyze SOP formation mechanisms by comparing the electrical, optical and spectral properties of positive and negative DC glow discharges interacting with a grounded water surface. Key differences in gas temperature, electric field and reactive species distribution are hence identified. For positive DC glow discharges (PGD), the gas temperature remains in the 350-370 K range, while the reduced electric field remains below 100 Td across the gap. The plasma is dominated by OH$^\bullet$ and N$_2$* species, whose excitation results from direct electron impact and energy transfer in a low-field environment. The absence of strong ionization and electric field gradients leads to a spatially homogeneous emission layer on the liquid surface, resulting in a circular uniform plasma (CUP) pattern without self-organization. In contrast, SOP emerge exclusively under negative DC glow discharges (NGD) at currents above 15 mA. These discharges are characterized by a non-linear reduced electric field, peaking at 485 Td at 1mm from the cathode pin, dropping below 100 Td in the central gap and rising to 460 Td near the water surface. There, the plasma layer still contains OH$^\bullet$ and N$_2$* species but also N$_2^+$ ions, the latter being critical for SOP formation. SOP morphology evolves with gap size: at 7 mm, patterns transition from specks to filaments, with pattern diameters and thickness as high as 5.5 mm and 210 µm respectively. Lowering water surface tension with surfactants reduces SOP size and modifies pattern morphology. Our results deepen understanding of plasma self-organization mechanisms, particularly the role of polarity and liquid surface dynamics. |
| Keywords | Positive DC glow discharge; Negative DC glow discharge; Self-organized patterns; liquid surface |

# I. Introduction

Self-organized patterns (SOP) of cold plasma refer to structured and recurring spatial arrangements that spontaneously emerge in plasma discharges under specific conditions [Trelles, 2016]. These patterns arise due to complex interactions between electric field, charged particles and reactive species within the plasma and manifest in various geometric forms such as rings, dots, filaments and more intricate designs [Zhu, 2014], [Trelles, 2016]a. Their formation and dynamics have been extensively studied due to their potential in divers applications such as surface modification, material processing, environmental remediation and plasma medicine [Foster, 2020], [Chen, 2020]. Their structured nature can enhance the spatial precision and controlled distribution of reactive species, facilitating more efficient pollutant degradation in environmental treatments and selective pathogen inactivation in plasma-based medical therapies.

At atmospheric pressure, the formation of SOP has been evidenced on material surface, such as on the counter-electrode of dielectric barrier devices (DBD) or on the surface of liquids exposed to DC glow discharges using single pin electrode devices (SPED).

In the case of DBD, self-organized filamentary structures appear with typical diameters in the range of 100–300 µm. Filament formation follows an activation-inhibition mechanism, where ionization from the electric field initiates the filament, while the charging of the dielectric barrier inhibits its radial diffusion [Callegari, 2014]. In this regime, small changes in parameters like voltage (e.g., 5–15 kV) or gas pressure (0.1–1 atm) can lead to significant bifurcations in filament patterns, resulting, for example, in hexagonal or spiral arrangements [Callegari, 2014], [Zhang, 2023]. Numerical and experimental studies have shown that external flow fields also significantly affect DBD filament dynamics. For instance, a flow of 1 m/s causes horizontal displacement of filaments, whereas higher flow velocities, such as 10 m/s, can induce dynamic patterns with directional filament motion [Zhang, 2023]. Additionally, filament patterns and structures can exhibit self-organized periodicity (SOP) due to memory effects. These effects arise from residual charges deposited on the dielectric surface during previous discharge cycles, which influence subsequent filament formation and behavior. Such residual charges, often concentrated at prior filament locations, modify the local electric field distribution in later cycles, leading to spatial regularity or shifts in filament positioning [Muller, 1999]. This memory-driven behavior contributes to structural complexity, such as the quincunx SOP, where high-current filaments are surrounded by lower-current discharges [Callegari, 2014]. SOP formation is further shaped by electric field redistribution caused by space charge effects and thermal instabilities induced by Joule heating. These mechanisms drive transitions between uniform and filamentary patterns, creating structures such as hexagons or rings [Raizer, 2013]. Variations in experimental conditions, including pressure or electrode geometry, play a crucial role in determining the dynamics and stability of these patterns.

As an alternative to DBDs, SPEDs can be employed to generate negative DC discharges on liquid surfaces at atmospheric pressure, enabling the formation of SOP. Alternatively, positive DC discharges result into the formation of continuous uniform patterns (CUP) which are inherently non-self-organized. The development of SOP is governed by key factors, including gap distance, discharge current, gas composition and liquid properties, as outlined below:

- The gap distance between the pin electrode and the liquid surface is a critical factor in determining SOP morphology. For values higher than 4.5 mm, SOP shift from conical and cylindrical forms to ring-like structures at the liquid surface, often with a homogeneous spot at the center [Diamond, 2020]. As the gap distance increases, the patterns become more defined. Chen *et al.* further emphasizes that shorter gap distances result in more intense and localized plasma interactions, leading to faster SOP formation, while larger distances reduce plasma intensity and produce broader and less concentrated structures [Chen, 2020].

- Another important factor is the discharge current. In atmospheric pressure glow discharges operating at low currents, small distinct spots appear, but as the current rises, these spots merge into larger and more complex structures [Verreycken, 2009]. This transition is observed across different electrolyte solutions, with currents ranging from 20 to 80 mA [Kovach, 2021]. Shirai *et al.* found that increasing discharge current, alongside factors like helium flow and gap length, leads to more stable and pronounced SOP at the liquid anode [Shirai, 2011].

- Electrolyte conductivity also plays a significant role in SOP stability and complexity. As conductivity increases, the discharge stabilizes, allowing rapid formation of patterns such as discs, rings and stars [Zhang, 2018], [Zheng, 2015]. Conductivities as high as 4820 µS result in immediate pattern formation upon discharge initiation [Ghimire, 2023]. Additionally, higher conductivities lead to fewer but more stable plasma dots with shorter propagation lengths at elevated voltages [Herrmann, 2023]. Shirai *et al.* also confirmed that electrolyte conductivity and concentration critically influence SOP stability, with variations in anode composition and helium flows leading to diverse organized structures [Shirai, 2009].

- The gas composition within the plasma-liquid interaction region further affects SOP formation. For instance, the presence of oxygen in the gap enhances pattern formation, producing more defined luminous spots or rings [Shirai, 2014]. As an electronegative gas, oxygen enables negative ion accumulation near the water surface, forming voltage-dependent loop patterns (single, triple, concentric) that emerge after bridging the electrode gap [Li, 2017].

- Liquid properties, including temperature and pH, are additional factors influencing SOP formation [Zhang, 2018]. Discharges involving a liquid anode show that SOP stability and complexity increase when the liquid's temperature rises and its pH shifts toward acidic levels [Srivastava, 2022]. This interaction between plasma-generated species and the liquid environment produces unique patterns dependent on liquid characteristics like evaporation rates and ionic strength [Srivastava, 2022].

- Instability mechanisms in arc discharges pose major challenges to SOP formation. The competition between forcing and dissipative factors are responsible for driving non-linear plasma dynamics, resulting in thermal instability and energy transfer between electrons and heavy species [Trelles, 2013].

Among the DC discharges that a SPED can generate in ambient air, one has to clearly distinguish corona discharges from glow discharges. A corona discharge occurs when the electric field around a conductor exceeds the ionization threshold of the surrounding gas, typically air, but remains below the level required for complete breakdown [Lowke, 2003]. This leads to localized ionization, producing a faint bluish glow and reactive species like ozone. The process begins with field concentration around a sharp conductor under high voltage, where the electric field surpasses the dielectric breakdown strength of air (~3 kV/mm). Free electrons accelerate and collide with neutral molecules, triggering an ionization cascade that forms a plasma region [Goldman, 1978]. As charges recombine, light is emitted, and chemical reactions occur in the surrounding environment. However, at higher currents, increased gas temperatures (typically > 1000 K) suggest a transition from classic corona to bulk ionization and thermal effects, characteristic of glow discharge, which operates at higher energy levels than traditional corona discharge [Staack, 2005]. Therefore, in this article, we refer either to negative DC discharges (NGD) or positive DC glow discharges (PGD).

To date, no study has directly compared positive and negative DC glow discharges to understand why SOP form exclusively in the latter. This article addresses this gap by offering a comprehensive analysis of the electrical, optical and spectral properties of both discharge types. Additionally, it examines the axial distribution of the reduced electric field and active species to elucidate the mechanisms by which cold plasma layers transform into CUP or SOP at the water surface.

# 2. Materials and Methods

## 2.1. Experimental setup

A DC glow discharge is generated in ambient air between a pin electrode and a grounded water sample, as sketched in **Figure 1**. The pin electrode (rod-shaped electrode of tungsten with a 3.2 mm diameter and a 25° half-angle tip) is polarized at high voltage, either positive or negative, using a DC generator (FuG Elecktronik GmbH company, HCN 350-6500 series). The liquid sample consists of 100 mL of tap water contained in a glass vessel, 8 cm in diameter. The gap is defined as the distance between the tip of the pin electrode and the water surface. The pin electrode is fixed to a linear translation stage so that the gap can be changed from 1 mm to 25 mm, with a vertical accuracy of ±10 µm. A ballast resistor (150 kΩ, 800 W) is placed between the DC power supply and the pin electrode to protect against short-circuit current in case of arcing.

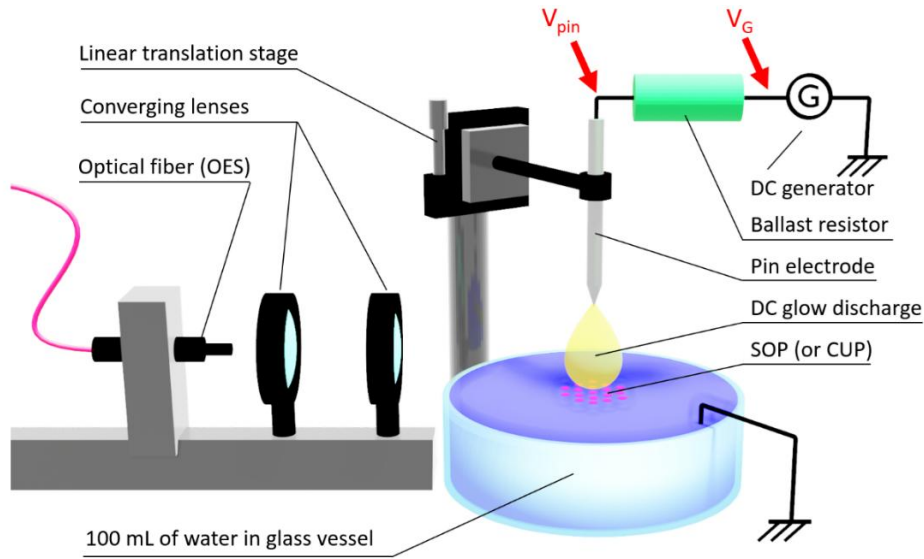

*Figure 1. Schematic of the single pin electrode device (SPED) generating a negative DC glow discharge (NGD) above water surface. Here, the layer of cold plasma formed on water surface is a self-organized pattern (SOP).*

## 2.2. Technique analyses & Measurements methods

### 2.2.1. Electrical characterizations

The discharge current ($I_d$) is calculated using Ohm's law by measuring the voltage drop across the ballast resistor (150 kΩ) as the difference between $V_{pin}$ and $V_G$, then dividing by the resistor value. Based on the principle of charge conservation, it is assumed that the discharge current ($I_d$) is equal to the ballast current ($I_b$). The discharge voltage ($V_d$) is approximated as the pin voltage ($V_{pin}$), since the water sample resistance, measured at approximately 65 Ω via LCR analysis, is negligible in this setup. Measurements of $I_d$ and $V_d$ are performed using a digital oscilloscope (Teledyne LeCroy Wavesurfer 3054), equipped with a high-voltage probe (Tektronix P6015A, 1000:1).

### 2.2.2. Camera imaging

Photographs of the DC glow discharge are captured using a Nikon camera (D610 model) equipped with a SIGMA lens (MACRO model, 105 mm of focal length, f/2.8 maximum aperture, EX DG OS HSM). Images are taken with an exposure time of 5 ms and an aperture setting of f/5.0, allowing a substantial amount of light while maintaining a moderate depth of field. To prevent sensor saturation, the following measures are employed: (i) real-time histogram monitoring, (ii) automatic ISO sensitivity control and (iii) camera's overexposure alert function. Spatial profiles, including height, diameter and thickness of DC glow discharges, CUP and SOP, are analyzed from the captured images using ImageJ software (version 1.54g).

### 2.2.3. Temperature measurements

Temperatures of gas, liquid and pin electrodes are measured combining two complementary approaches: infrared imaging and optical emission spectroscopy.

Thermal mapping of the SPED is carried out using an infrared camera (Jenoptik & Infratec VarioCAM® HD head 680/30) completed with a close-up lens (Converter Jenoptik ×0.2) hence ensuring pixel resolution of 76 μm at 70 mm focus distance. The temperature of the probed surfaces is recorded based on their infrared emission properties. To ensure accurate temperature measurements, the emissivity coefficient (ε) is set to 0.96 for the water surface and 0.35 for the tungsten pin electrode. The actual temperature ($T_{IR,act}$) is related to the measured temperature ($T_{IR,meas}$) by Equation (1):

$$T_{IR,act} = \frac{T_{IR,meas}}{\varepsilon} \tag{1}$$

Optical emission spectroscopy (OES) is carried out using an Andor SR-750-B1-R spectrometer equipped with an ICCD camera (model Istar) in a Czerny-Turner configuration. The system has a focal length of 750 mm and a 1200 groove/mm grating, blazed at 500 nm. For all experiments, parameters include an exposure time of 0.4 s, 15 accumulations and a gain level of only 1. Sep'n'glue procedures are performed to obtain a spectrum over large spectral band, i.e. from 250 to 800 nm. Axial profiles along the gap are obtained with a 1 mm accuracy using an OES fiber coupled with two converging lenses (focal lengths of 60 and 100 mm), as sketched in **Figure 1**.

The rotational temperature of the DC glow discharge (i.e. cold plasma of ambient air) is determined by analyzing the second positive system (SPS) of molecular nitrogen ($N_2$) emissions, specifically the ($C^3\Pi_u \rightarrow B^3\Pi_g$) transitions. The collected spectra are calibrated for wavelength and intensity using a deuterium lamp, ensuring high precision in the analysis. The rotational temperature is obtained by fitting the experimental emission spectra to simulated spectra based on the Boltzmann distribution of rotational states. The simulation relies on the molecular constants of molecular nitrogen, its spectral transitions (P, Q and R branches) and incorporates Hönl-London coefficients to account for rotational line intensities. This fitting process is performed using a custom-developed algorithm, which minimizes the difference between the measured and simulated intensity distributions. The simulations consider the spectral resolution of the Andor spectrometer (50 pm) and account for potential broadening effects.

Once the gas temperature is determined, the gas density is deduced using Equation (2).

## 2.2.4. Reduced electric field measurements

The reduced electric field in the interelectrode region of the SPED is measured using the method proposed by Paris, which is based on the intensity ratio of two nitrogen spectral bands measured in a cold plasma of dry air [Paris, 2005]. Foster et al. applied a similar method for DC glows interacting with a liquid anode, utilizing the $N_2^*$ band at 391.4 nm (first negative system) and the $N_2^*$ band at 394.3 nm (second positive system) [Foster, 2020]. In the present study, due to the low detectability of the 394.3 nm band, the analysis is carried out on the more prominent bands at 337.1 nm and 391.4 nm to ensure greater accuracy. The intensity ratio of these bands is defined by Equation (3) and is labelled $R_{dry}$ to remind that the calibrations curves provided by Paris et al. are only relevant for a cold plasma of dry air. $R_{dry}$ provides the basis for calculating the reduced electric field using Equation (4), which is derived from *Fig. 6* in [Paris, 2005]. The reduced electric field is expressed in Townsends (Td), corresponding to $10^{-21}$ V·m$^2$.

$$n = \frac{P}{k_B T} \tag{2}$$

$$R_{dry} = \frac{I_{391}^{dry}}{I_{337}^{dry}} \tag{3}$$

$$R_{dry} = 2.429 \times \left(\frac{E}{n}\right)^{-55.34 \times \left(\frac{E}{n}\right)^{-0.724}} \tag{4}$$

To refine the method of Paris in a context of ambient air which contains water vapor, quenching effects have been considered. For a given excited species X standing for $N_2(C^3\Pi_u)$ or $N_2(B^2\Sigma_u^+)$, the total de-excitation rate is provided by Equation (5) where $A_X$ is the spontaneous emission rate, $Q_X$ is the collisional quenching rate and $N_X$ is the population of the excited state. The quenching rate is given by Equation (6) where $k_{N_2 \to X}$, $k_{O_2 \to X}$ and $k_{H_2O \to X}$ are the quenching rates coefficient of X respectively induced by $N_2$, $O_2$ and $H_2O$. $[N_2]$ and $[H_2O]$ are the number densities of nitrogen and water vapor expressed in cm$^{-3}$. Since the observed emission intensity of a spectral line at wavelength $\lambda_X$ is given by Equation (7) where the steady-state population of X is proportional to the inverse of the total de-excitation rate (Equation (8)), it turns out that this line intensity can be expressed by Equation (9) where the species X is quenched by $N_2$, $O_2$ and water vapor in ambient air.

$$\frac{dN_X}{dt} = -(A_X + Q_X).N_X \tag{5}$$

$$Q_X = k_{N_2 \to X}[N_2] + k_{O_2 \to X}[O_2] + k_{H_2O \to X}[H_2O] \tag{6}$$

$$I_X \propto A_X.N_X \tag{7}$$

$$N_X \propto \frac{1}{A_X + Q_X} \tag{8}$$

$$I_X^{wet} \propto \frac{A_X}{A_X + k_{N_2 \to X}[N_2] + k_{O_2 \to X}[O_2] + k_{H_2O \to X}[H_2O]} \tag{9}$$

$$I_{337}^{wet} \propto \frac{A_{337}}{k_{N_2 \to N_2^*}[N_2] + k_{O_2 \to N_2^*}[O_2] + k_{H_2O \to N_2^*}[H_2O]} \tag{10}$$

$$I_{391}^{wet} \propto \frac{A_{391}}{k_{N_2 \to N_2^+}[N_2] + k_{O_2 \to N_2^+}[O_2] + k_{H_2O \to N_2^+}[H_2O]} \tag{11}$$

$$R([H_2O]) = \frac{I_{391}[H_2O]}{I_{337}[H_2O]}$$
$$= \frac{A_{391}}{A_{337}} \cdot \frac{k_{N_2 \to N_2^*}[N_2] + k_{O_2 \to N_2^*}[O_2] + k_{H_2O \to N_2^*}[H_2O]}{k_{N_2 \to N_2^+}[N_2] + k_{O_2 \to N_2^+}[O_2] + k_{H_2O \to N_2^+}[H_2O]} \tag{12}$$

$$\frac{A_{391}}{A_{337}} = R_{dry} \cdot \frac{k_{N_2 \to N_2^+}[N_2] + k_{O_2 \to N_2^+}[O_2]}{k_{N_2 \to N_2^*}[N_2] + k_{O_2 \to N_2^*}[O_2]} \tag{13}$$

$$\underbrace{R_{dry}}_{unknown} = \underbrace{R([H_2O])}_{measured} \cdot \frac{1 + \frac{k_{H_2O \to N_2^+}[H_2O]}{k_{N_2 \to N_2^+}[N_2] + k_{O_2 \to N_2^+}[O_2]}}{1 + \frac{k_{H_2O \to N_2^*}[H_2O]}{k_{N_2 \to N_2^*}[N_2] + k_{O_2 \to N_2^*}[O_2]}} \tag{14}$$

At atmospheric pressure, collisional quenching dominates over spontaneous emission ($A_X \ll Q_X$), because excited species undergo frequent collisions before emitting photons. Applying this to the two spectral lines of interest leads to the expressions of (10) and (11) for $N_2(C^3\Pi_u)$ at 337.1 nm and $N_2(B^2\Sigma_u^+)$ at 391.4 nm respectively.

The intensity ratio of these lines, expressed in Equation (12), provides a reference case corresponding to dry air (i.e., without water vapor), given by Equation (13). Substituting (13) in (12) yields (14) where $R([H_2O])$ is the line ratio measured in this work and $R_{dry}$ is the reference case of Paris (dry air) which remains unknown and must be determined to apply the method of Paris [Paris, 2005].

The values of the quenching rate coefficients used in this study are given in **Table 1** while the values of $[N_2]$, $[O_2]$ and $[H_2O]$ are expressed in molecules.cm$^{-3}$ and determined by quadrupole mass spectrometry (Model HPR-20, Hiden Analytical Ltd). Specifically, the absolute values of $[N_2]$ and $[O_2]$ are obtained by calibrating the mass spectrometer (MS) with two reference gas cylinders from Air Liquide company, each containing 500 mol ppm of $N_2$ and $O_2$, respectively. The concentration of $H_2O$ is measured in two steps. First, in a controlled air environment, humidity levels of 0%, 20%, 50%, and 90% are set and monitored using a hygrometer, while the corresponding $H_2O$ intensity is recorded using MS. Second, in the DC glow discharge, the intensity of $H_2O$ is measured directly. The recorded $I_{MS}$ intensities are then converted into relative pressures and ultimately expressed in molecules/cm³.

**Table 1. Quenching rate coefficients**

| Notations | Values | References |
|---|---|---|
| $k_{H_2O \to N_2^*}$ | $7.1 \ 10^{-10} \ cm^3.s^{-1}$ | [Morozov, 2005] |
| $k_{N_2 \to N_2^*}$ | $2.5 \ 10^{-11} \ cm^3.s^{-1}$ | [Bak, 2011] |
| $k_{O_2 \to N_2^*}$ | $3.0 \ 10^{-10} \ cm^3.s^{-1}$ | [Pancheshnyi, 2000] |
| $k_{H_2O \to N_2^+}$ | $8.6 \ 10^{-10} \ cm^3.s^{-1}$ | [Pancheshnyi, 1998] |
| $k_{N_2 \to N_2^+}$ | $8.2 \ 10^{-10} \ cm^3.s^{-1}$ | [Jolly, 1983] |
| $k_{O_2 \to N_2^+}$ | $5.1 \ 10^{-10} \ cm^3.s^{-1}$ | [Pancheshnyi, 1998] |

### 2.2.5. Surface tension measurements

Surface tension of water is measured following a protocol based on the Tate's law. First, the water is injected into a capillary tube with a 1.25 mm radius. 10 drops of the liquid are weighted on a micro-analytical balance so as to accurately determine the average mass of a single droplet ($m_{drop}$). Two forces act on this droplet: the weight which, following equation (15) is the product of $m_{drop}$ by the gravity constant (g) and the capillary force given by equation (16) where $\gamma$ denotes the surface tension and $r_{cap}$ the internal radius of the capillary. At the point of detachment, the droplet's weight equals the capillary force, leading to the equilibrium in equation (17). From this, the Tate's law formula is derived to calculate the surface tension in equation (18).

$$W_{drop} = m_{drop} \times g \qquad (15)$$

$$F_{cap} = 2\pi \times r_{cap} \times \gamma \qquad (16)$$

$$W_{drop} = F_{capil} \qquad (17)$$

$$\gamma = \frac{m_{drop} \times g}{2\pi \times r_{cap}} \qquad (18)$$

# 3. Results

## 3.1. Electrical, optical and spectral properties of negative and positive DC glow discharges

### 3.1.1. Electrical characterization

First, the influence of the DC voltage — positive or negative —is studied on the electrical and optical properties of the discharge. In each configuration, the water sample is always grounded.

In the pin cathode configuration, NGD is characterized by a V-I curve (**Figure 2a**) showing a breakdown voltage of −1.5 kV at −8 mA. As the current increases, the operating voltage slightly decreases and stabilizes around −1.3 kV. In the same current range, the electrical power rises linearly from 12 to 42 W (**Figure 2b**). Notably, SOP only appear within a specific range (highlighted in the blue window). As illustrated in **Figure 2c**, the electric field is directed toward the pin electrode, driving primary electrons toward the liquid surface, which is polarized as the anode. Consequently, two distinct emissive regions can be distinguished in **Figure 2c**: (i) the NGD which takes on an obovate profile with diminishing light intensity from the pin electrode; and (ii) the cold plasma layer which appears as a regularly structured sub-millimeter pattern, i.e. a SOP at the liquid surface.

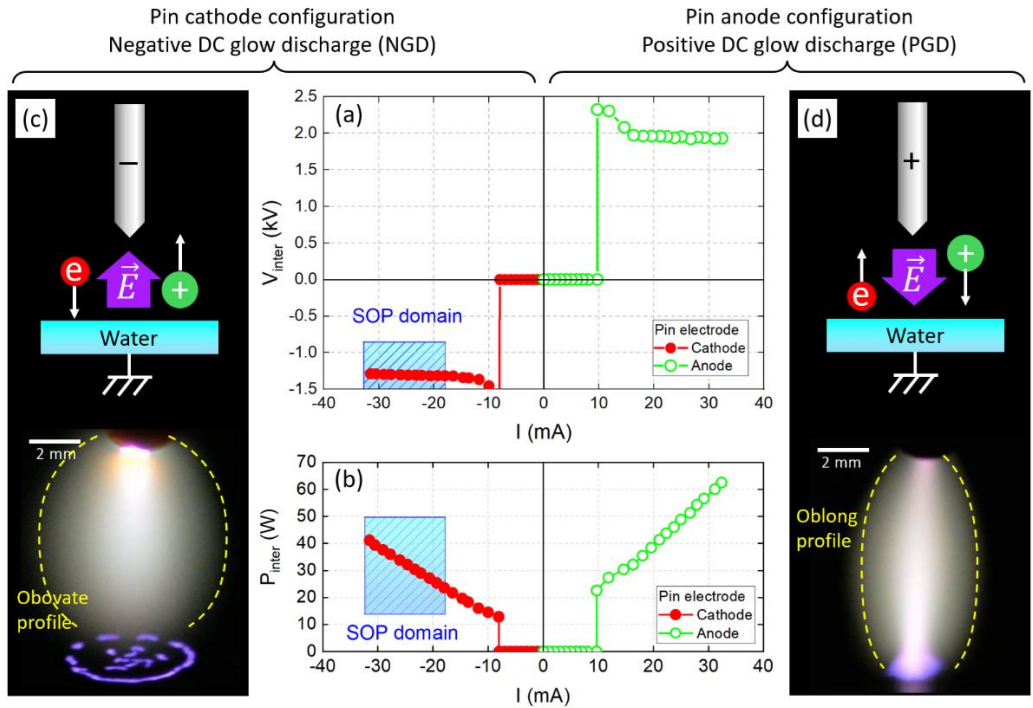

**Figure 2.** Discharge characteristics with the pin electrode configured as either a cathode (negative DC glow discharge, NGD) or an anode (positive DC glow discharge, PGD). (a) Voltage-current curves for both configurations, showing the operating ranges of the discharge, (b) Power-current curves, (c) Schematics and photograph of obovate-profile NGD and SOP at water surface (d) Schematics and photograph of oblong-profile PGD and CUP at water surface. All photographs and plots are obtained for a 8 mm gap.

When the pin electrode is polarized as an anode, the positive DC glow discharge (PGD) presents a V-I curve with a breakdown voltage of +2.3 kV at 10 mA, stabilizing at 1.9 kV for higher currents (**Figure 2a**), while the electrical power increases from 22 to 63 W (**Figure 2b**). In this configuration, the electric field is oriented toward the water surface and two new distinct emissive regions appear in **Figure 2d**: (i) the PGD which, contrarily to the previous configuration, exhibits now an oblong profile that bridges the pin electrode and the water surface, with strong and continuous light intensity throughout the gap; and (ii) the cold plasma layer at the water surface forming a circular uniform pattern (CUP), without any detectable self-organization.

The negative DC glow discharge operates at lower breakdown voltage (−1.6 kV) than the positive (+2.3 kV), with SOP forming only in the negative configuration when the current exceeds 15 mA. This shows that polarity directly influences self-organization and that even higher power in the positive glow discharge does not induce SOP. Understanding how polarity affects charged species densities and plasma reactions is crucial to understand SOP formation mechanisms, as discussed in Section 4.

### 3.1.2. Light intensity profiles

As clearly evidenced in both anode (**Figure 3a**) and cathode configurations (**Figure 3b**), the most emissive regions of DC glow discharges in ambient air correspond to areas of higher excitation and ionization, driven by locally intensified electric field.

In the pin anode configuration (**Figure 3a**), the light emission of PGD is the more intense between 2 and 5 mm above the water surface, hence indicating a region of strong excitation. Conversely, the pin cathode configuration in **Figure 3b** shows highest light intensity for z = 0-2 mm, where electric field is assumed to reach its maximum value. Beyond this region, the light intensity of NGD decreases sharply with increasing z-values.

These distinct light intensity profiles highlight the excitation regions that are specific to each pin polarization, potentially involving different reaction mechanisms and different active species densities, as discussed in Section 4.

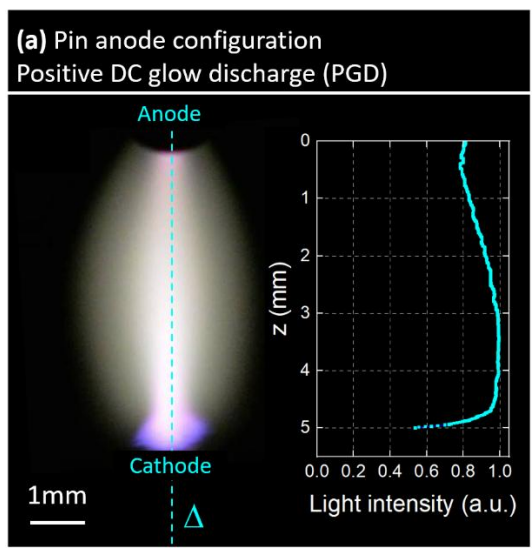

**(a)** Pin anode configuration
Positive DC glow discharge (PGD)

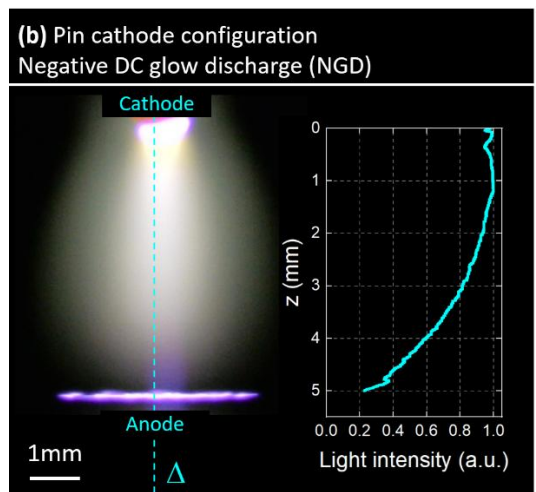

**(b)** Pin cathode configuration
Negative DC glow discharge (NGD)

*Figure 3. Photographs and light intensity profiles of the DC glow discharge (a) in the pin anode configuration where CUP are formed and (b) in the pin cathode configuration where SOP can be formed. Light intensity profiles are plotted by measuring light exclusively along the Δ vertical line. In both configurations: gap = 5 mm, $I_d$ = −25 mA, $V_d$ = − 1.3 kV.*

### 3.1.3. Spectral and axial distribution of emissive species (gap = 8 mm)

To further validate the previous hypotheses, the plasma phase confined in the 8-mm gap is characterized by OES. Axial measurements are performed at 1-mm intervals, covering a spectral range of 250–800 nm for both the PGD (**Figure 4**) and NGD (**Figure 5**) configurations.

In the pin anode configuration, PGD exhibits strong emission lines and bands, particularly below 400 nm, as shown in **Figure 4a**. Several observations are noteworthy:

- The band headed at 308.9 nm corresponds to the presence of $OH^\bullet$ radicals generated throughout the gap. As shown in **Figure 4b**, the intensity of the $OH^\bullet$ emission increases near the liquid surface (z = 7.5 mm), likely due to the higher concentration of water vapor in this region, enhancing electron-impact dissociation. At 337.1 nm (**Figure 4c**), the Second Positive System of $N_2$ is most prominent near the high-voltage electrode ($I_{opt}$ = 40 a.u.). Then, it drops significantly between 2.5-6.5 mm ($I_{opt}$ < 5 a.u.) before slightly rising again in the plasma layer near the liquid surface ($I_{opt}$ = 10 a.u.). This behavior suggests differing ionization and excitation dynamics across the gap.
- In the 388-392 nm range (**Figure 4d**), a spectral band peaking at 389.5 nm is detected, corresponding to excited molecular nitrogen ($N_2^*$, SPS). However, no peak is observed at 391.4 nm, which would have indicated the presence of $N_2^+$ ions.
- Oxygen emission line at 777.4 nm is detected (**Figure 4e**), though its amplitude is quite low (around 0.5 a.u.), making its interpretation more speculative.

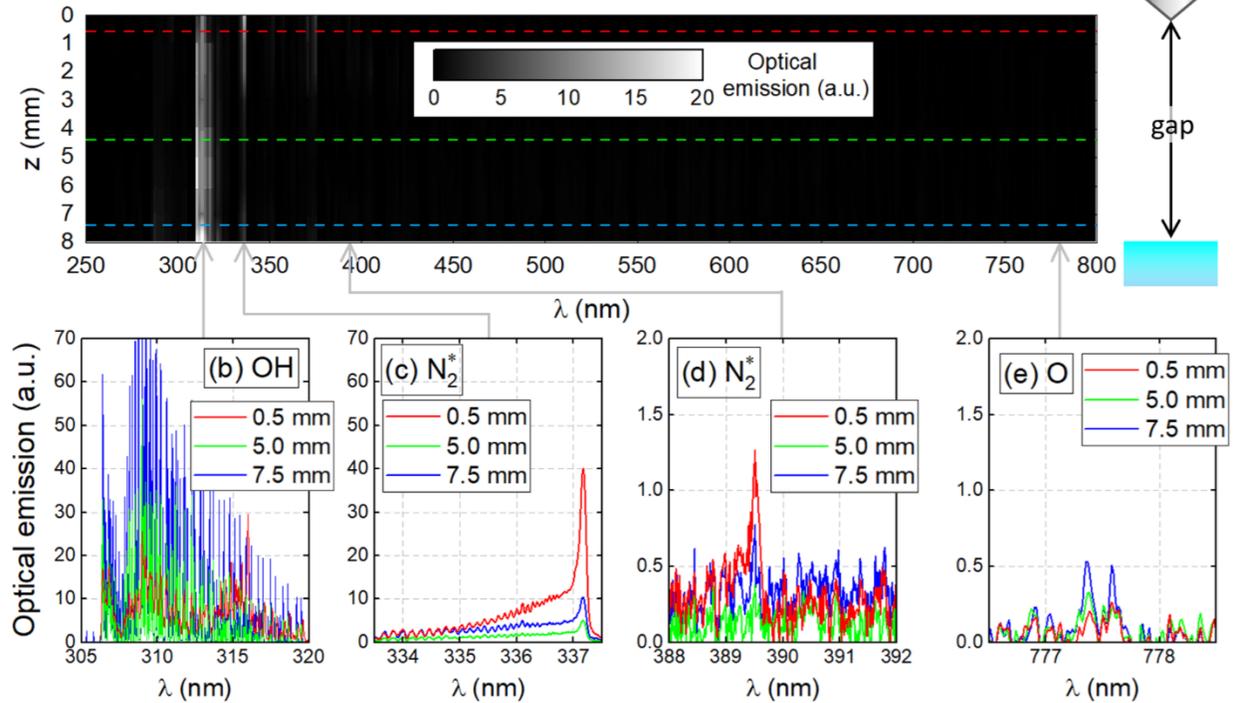

*Figure 4. Pin anode configuration, $I_d = -25$ mA, $V_d = -1.3$ kV. (a) z-spectral diagram obtained by optical emission spectroscopy for gap = 8 mm. (b-e) Optical emission spectra obtained at 3 axial locations within the gap, namely 0.5 mm, 5.0 mm and 7.5 mm. (b) OH* band, (c-d) $N_2$ bands, (e) O* line.*

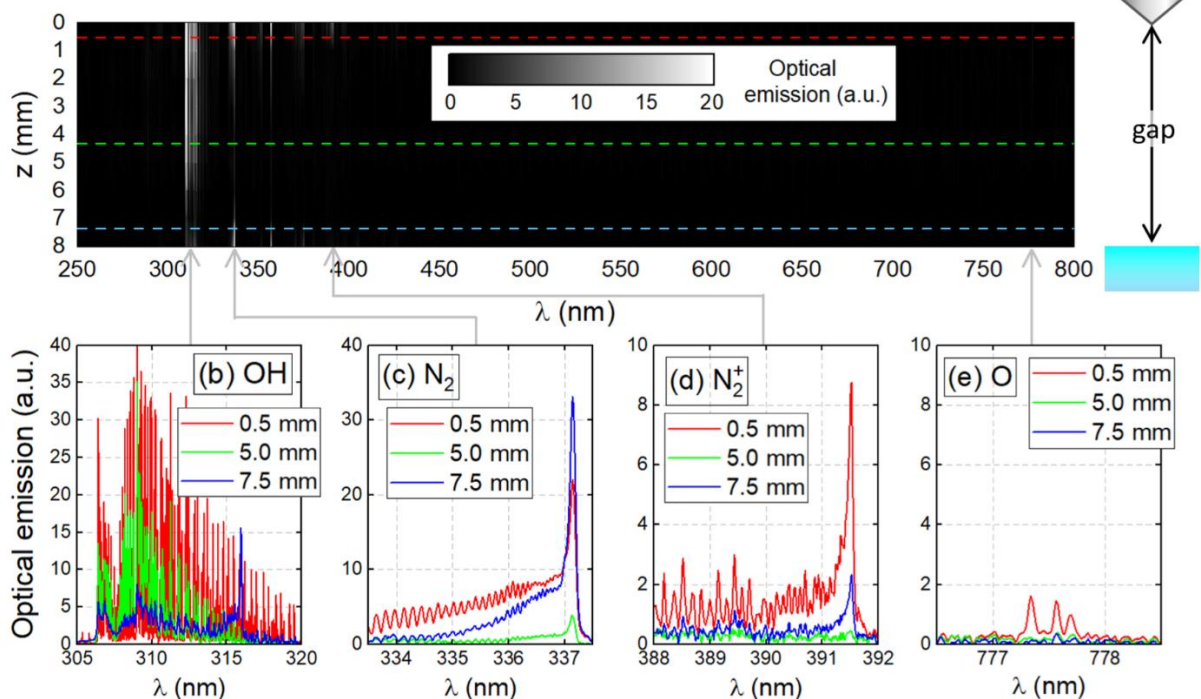

*Figure 5. Pin cathode configuration, $I_d = -25$ mA, $V_d = -1.3$ kV. (a) z-spectral diagram obtained by optical emission spectroscopy for gap = 8 mm. (b-e) Optical emission spectra obtained at 3 axial locations within the gap, namely 0.5 mm, 5.0 mm and 7.5 mm. (b) OH* band, (c) $N_2$ band, (d) $N_2^+$ band, (e) O* line.*

In the pin cathode configuration, electron generation predominantly occurs at the pin electrode rather than in the central region of the gap. As shown in **Figure 5a**, this leads to a highly intense glow localized in the vicinity of the pin electrode while NGD gradually vanishes for increasing z-values, until SOP formation on the liquid surface. Consequently, the spatial distribution of gaseous emissive species in this configuration (**Figure 5a**) differs markedly from that observed in the anode configuration (**Figure 4a**). Here again, several observations are noteworthy:

- The $OH^\bullet$ emission bands (**Figure 5b**) are significantly more intense near the pin electrode than at the water surface. This contrasts with the pin anode configuration (**Figure 4b**), where weak $OH^\bullet$ emission near the pin suggested insufficient electron energy for $OH^\bullet$ formation. Despite the higher water vapor density near the water surface, efficient $OH^\bullet$ production is not observed in this configuration.
- The molecular nitrogen bands ($N_2$, SPS) (**Figure 5c**) exhibit the same trends as in the anode configuration: strong emission near the pin electrode, weak emission in the central gap and a slight increase in intensity near the water surface. However, the pin cathode configuration differs significantly in that $N_2^+$ ions are detected at 391.4 nm (**Figure 5d**), visible both near the pin electrode and at the water surface.
- Oxygen radicals are detected near the pin electrode, with an intensity below 2 a.u. (**Figure 5e**). This differs from the pin anode configuration (**Figure 4e**), where oxygen emission was stronger near the water surface. This result suggests that radiative oxygen production is closely tied to the formation of $OH^\bullet$ radicals, which are more prevalent near the cathode.

### 3.1.4. Axial distribution of gas temperature

Before analyzing the optical emission spectra (OES) of $N_2^*$ (**Figure 4c** and **Figure 5c**) to determine the gas temperature in PGD and NGD, an infrared thermographic camera is employed to measure the temperatures of both the pin electrode and the water surface within the experimental setup. **Figure 6a** presents the thermal mapping of the electrode in both pin anode and pin cathode configurations. In the PGD configuration, the electrode temperature exhibits a gradient, increasing from 480 K to 615 K as one approaches the pointed extremity. In contrast, **Figure 6a** also shows that in the pin cathode configuration (NGD), the electrode temperature remains uniform and slightly higher, estimated at 640 K. In both cases, the water surface temperature in contact with the DC glow discharge remains comparable, ranging between 420 K and 450 K (**Figure 6b**).

The axial temperature profiles measured by infrared thermography ($T_{IR}$) are depicted in **Figures 6c** and **6d** for the positively and negatively polarized SPED, respectively. In these figures, the solid blue bars represent unambiguous temperature values, i.e. those of the electrode tip and water surface which are determined considering their respective emissivity coefficients ($\varepsilon_{H2O}$ = 0.96, $\varepsilon_{Tungst}$ = 0.35) . Conversely, the hatched bars correspond to the temperature of the DC glow discharge itself, whose interpretation is more speculative due to the infrared transparency of air molecules at most wavelengths. However, water vapor is known to absorb and emit radiation in specific infrared spectral bands. Given that both PGD and NGD exhibit localized regions of strong optical absorption (**Figure 3**) and RH levels as high as 94 %, the $T_{IR}$-derived gas temperatures appear consistent with expected values.

To further investigate gas temperature distribution, an alternative approach involves optical emission spectroscopy (OES) to determine the rotational temperature of the DC glow discharge by analyzing the rovibrational structure of the $N_2$ second positive system (SPS) bands. The resulting temperature profiles for NGD and PGD are presented in **Figures 6e** and **6f**, respectively. Additionally, the optical emission of $OH^\bullet$ radicals is included to highlight the regions where $H_2O$ molecules undergo the most significant dissociation. For increasing z-values, the gas temperature of PGD raises from 625 K (at pin's tip) to over 2000 K (on water surface), whereas the NGD gas temperature decreases from approximately 2350 K to 450 K.

Overall, these temperature estimates appear overestimated, particularly in the case of PGD. A water sample exposed to temperatures as high as 2350 K would have fully evaporated within minutes, a phenomenon never observed in our experimental setup. This overestimation arises from a limitation of the OES method, as detailed in Section 2.2.3. Specifically, the theoretical spectrum of $N_2^+$ does not account for quenching effects due to collisions with $N_2$, $O_2$, and $H_2O$ molecules. As a result, the hatched bars in **Figures 6e** and **6f** remain speculative. In contrast, the solid red bars, which correspond to regions characterized by low relative humidity and minimal $OH^\bullet$ emission, are considered reliable and closely align with infrared thermography data.

### 3.1.5. Diffuse nature of the DC glow discharges

To determine whether PGD or NGD could have exhibited short-lived streamers, fast ICCD imaging was performed with a 5 ns time resolution. However, no streamer structures were detected, and the discharge remained diffuse throughout the exposure. Additionally, DC current measurements were analyzed over time, revealing a stable, continuous signal without any peaks, even at small time scales. Such peaks would typically indicate transient ionization waves or streamer formation. The absence of these features in both optical and electrical diagnostics suggests that, under our experimental conditions, the discharge remains uniform and diffuse, rather than transitioning into a filamentary regime.

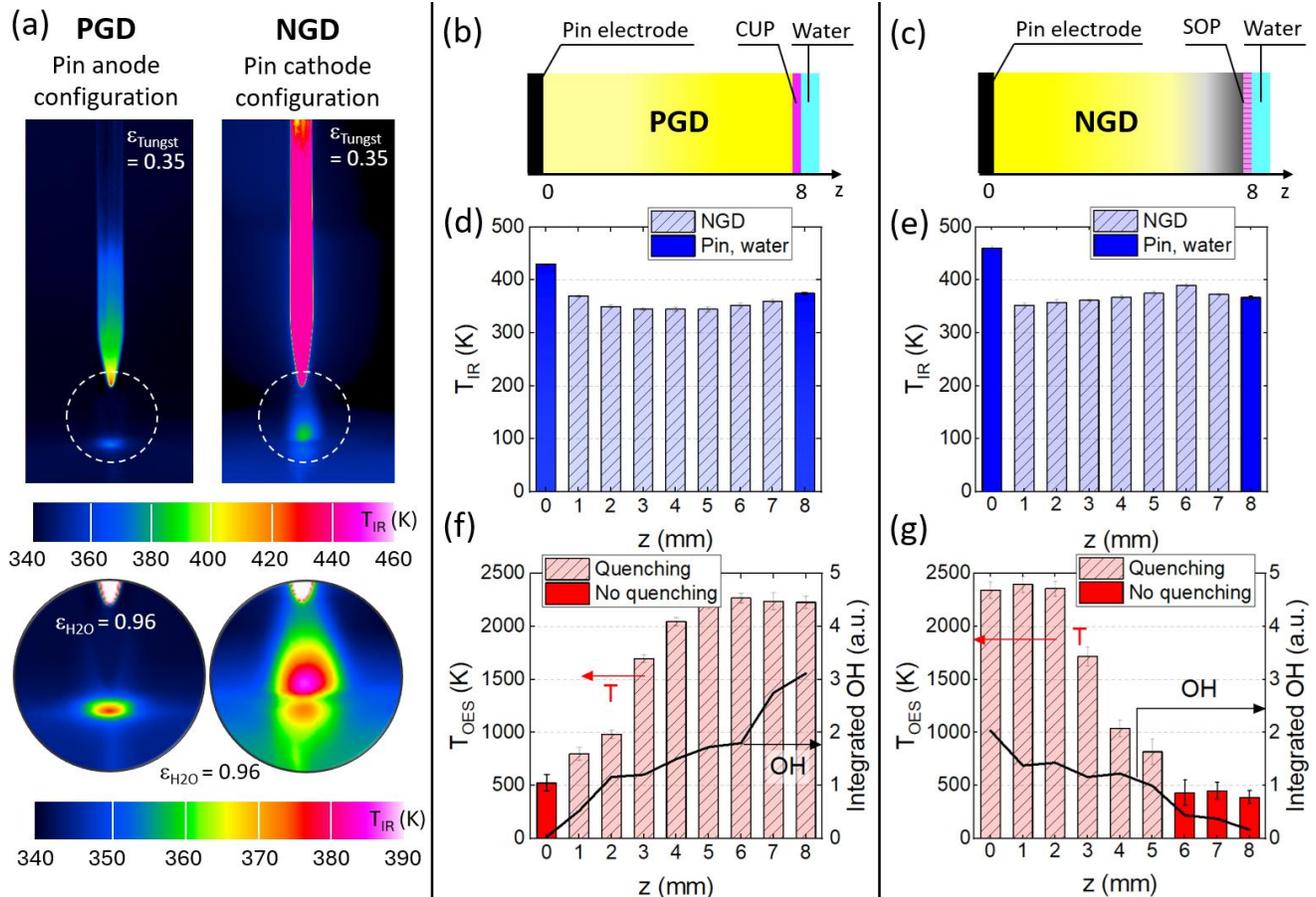

*Figure 6. (a) Infrared thermal map of pin electrode polarized positively (PGD) or negatively (NGD) above liquid surface. (b) Infrared thermal map of PGD or NGD in contact with the liquid surface. (c, d) Gas temperature profiles measured by IR camera for PGD and NGD respectively. (g, h) Gas temperature profiles measured by OES. The black curve represents the integrated value of the OH* band at 309 nm obtained by OES. $I_d = -25$ mA, $V_d = -1.3$ kV.*

## 3.2. Negative DC glow discharge: influence of physico-chemical parameters to modulate SOP properties

While both PGD and NGD have been characterized in the previous sections, the following analysis focuses exclusively on NGD. This choice is motivated by the fact that SOPs emerge only under negative polarization, whereas PGD results in a CUP at the liquid surface. Since the primary objective of this study is to investigate SOP formation mechanisms, we specifically analyze the influence of key parameters such as time, gap size, electric field, and surface tension in the NGD configuration.

### 3.2.1. Influence of time

SOP stability in the pin cathode configuration is studied in **Figure 7a** through a series of photographs captured at 5-minute intervals over a 20-minute period. The pin electrode is polarized at −1300 V with a current of −25 mA and a 8-mm gap. Under these conditions, the SOP primarily consist of specks and filaments. Specks are low aspect ratio elements resembling dots or short segments, approximately 500 µm in size and arranged in concentric patterns, as evidenced in the images at t = 10 min and t = 20 min. This discrete distribution remains consistent over time, with occasional clustering into higher-aspect-ratio elements referred to as filaments. These filaments exhibit various shapes, including curved (t = 5 min) and spiraled (t = 20 min) forms.

The transition from specks to filaments does not correlate with NGD operation time or the linear increase in water evaporation, which ranges from 0 to 5.4 mg over the 20-minute period (**Figure 7b**), with an evaporation rate of 0.27 mg/min. This indicates that SOPs dynamically reorganize in response to subtle environmental factors rather than major shifts in liquid properties. Since temporal factors alone cannot fully explain the speck-to-filament transitions, the influence of a more easily controlled variable – the gap – has been investigated in the next section.

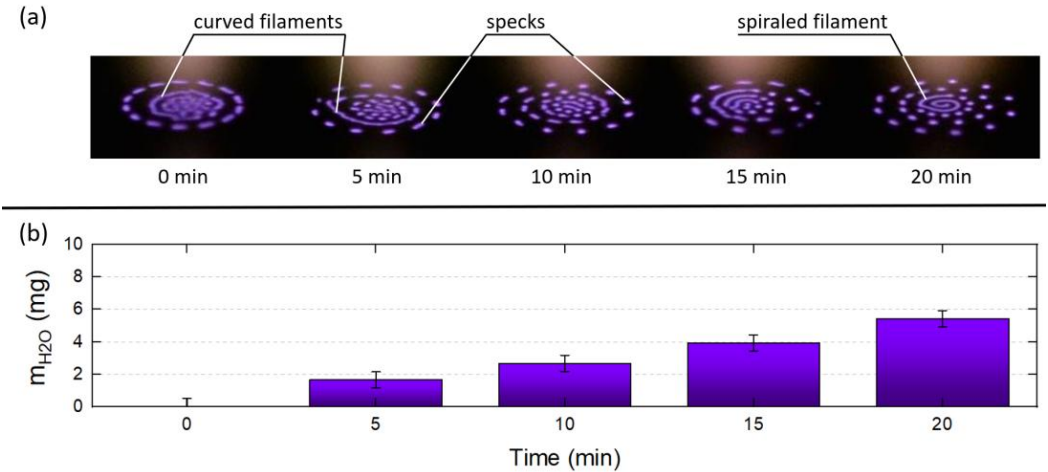

Figure 7. (a) Temporal evolution of SOP at the water surface under pin cathode configuration ($I_d$ = −25 mA, $V_d$ = −1300 V, gap = 8 mm). Photographs are taken every 5 minutes over a 20-minute period, showing transition between specks (low aspect ratio elements) and filaments (higher aspect ratio elements). (b) Water evaporation rate measured over the same 20-minute period.

### 3.2.2. Influence of gap

By increasing the gap between the pin electrode and the water surface, significant changes occur in both the electrical properties of the discharge and the morphology of the plasma layer. For gap values increasing from 1 to 12 mm, **Figure 8** indicates a voltage increase from approximately |−800| V to |−1600| V, accompanied by a current decrease from |−35| mA to |−29| mA. This corresponds to a voltage rise of 700 V and a current reduction of 6 mA, this resulting into a power dissipated by the DC glow discharge ($P_{d,max}$) increasing from 25 W to nearly 50 W. Beyond a 12-mm gap, however, the glow discharge becomes too unstable for accurate measurements.

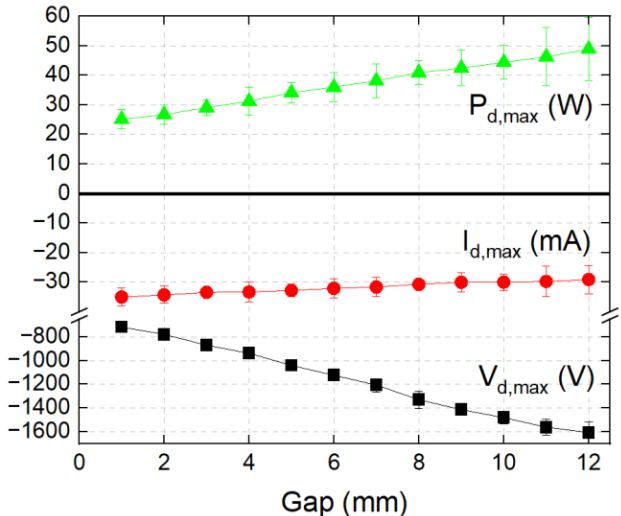

Figure 8. Influence of gap on the electrical parameters of NGD in pin cathode configuration. $I_{d,max}$ and $V_{d,max}$ represent the maximum current and voltage achievable with the DC high voltage power supply, respectively. $P_{d,max}$ is calculated as the product of $I_{d,max}$ and $V_{d,max}$.

In parallel with these observed electrical changes, significant morphological transformations of the plasma layer are evidenced in **Figure 9** when changing the gap. For values of 1-2 mm, the plasma layer forms a CUP with no discernible self-organization. Then, as the gap increases to 3 mm, the edges of the CUP begin to show signs of crenellation, signaling the initial onset of SOP. Further widening the gap to 4, 5 and 6 mm leads to the emergence of specks while at 7 mm, these specks evolve into filaments, which adopt curved shapes with characteristic lengths of about 6 mm. As the gap continues to expand to 8 and 9 mm, the self-organization becomes more pronounced, with filaments becoming more dominant as the number of specks decreases. The filaments elongate and organize into complete circular structures at a gap of 10 mm, exhibiting a high degree of self-organization. However, when the gap exceeds 12 mm, the SOP structures begin to disintegrate, and the plasma layer transitions back to a CUP-dominated morphology.

To clearly assess the extent to which the gap influences the characteristic dimensions of the plasma layer, its thickness is measured by capturing images of NGD perpendicular to the pin electrode. From **Figure 10a**, a light intensity profile is extracted, and the full width at half maximum (FWHM) is measured to quantify the SOP thickness, as illustrated in **Figure 10b**. The results, presented in **Figure 10c**, indicate that the CUP maintains a consistent thickness of approximately 140 µm for gap lower than 4 mm, whereas its thickness is increased to approximately 210 µm for gaps comprised between 4 and 10 mm. This broader thickness is indicative of the more complex structural organization associated with SOP formation. Beyond 10 mm, the SOP becomes too instable to obtain reliable measurements.

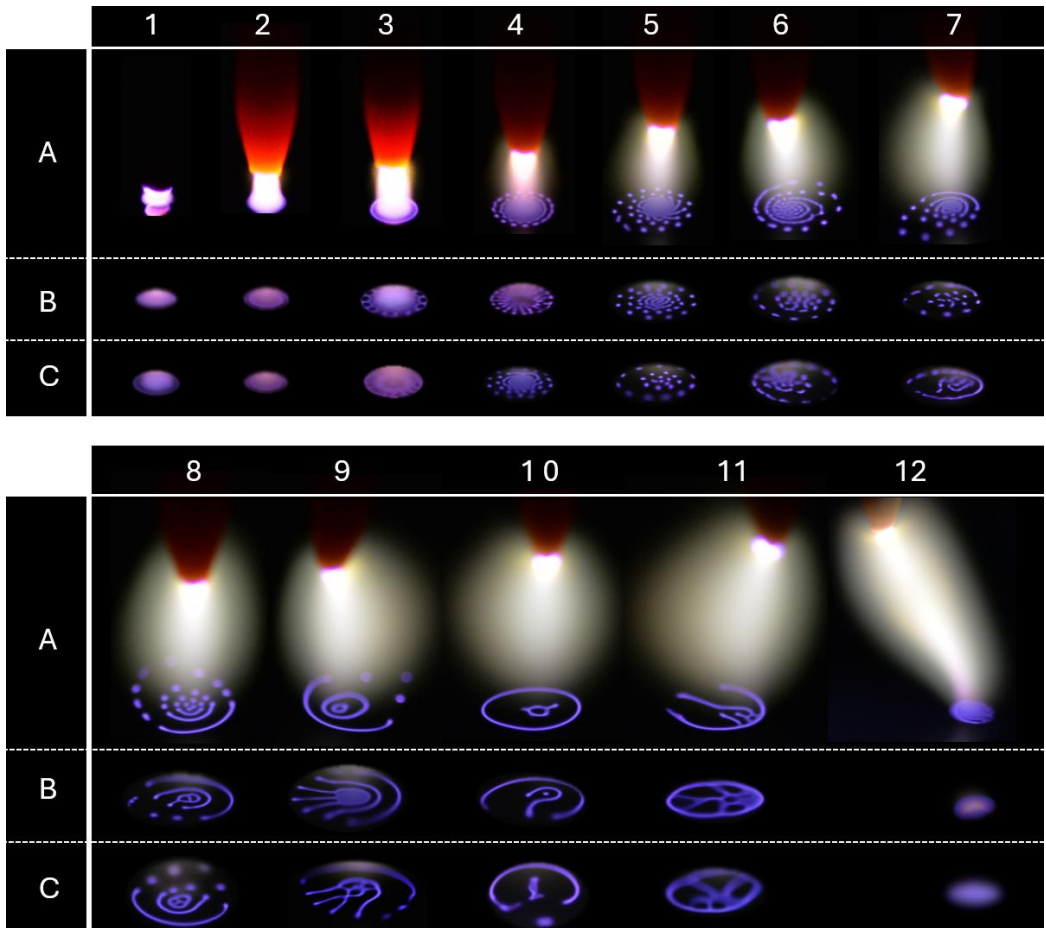

*Figure 9. Effect of varying the gap distance (from 1 mm to 12 mm) on the plasma layer formed at the water surface, showing the transition from a CUP to a SOP. The progression includes a transition from dot-like to filamentary structures, which eventually vanish as the gap increases. Rows A, B and C correspond to replicates under the same gap conditions.*

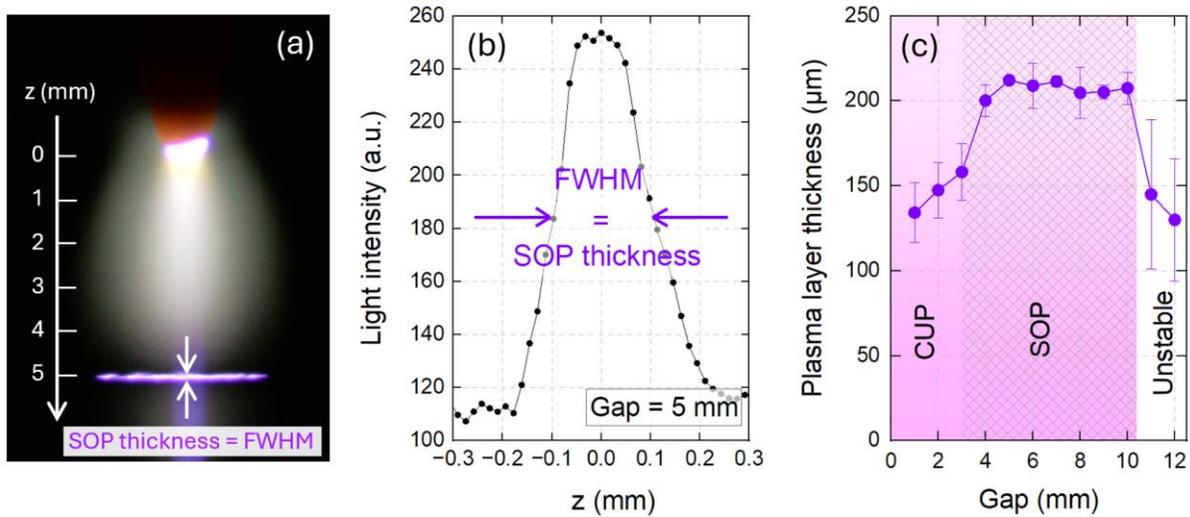

*Figure 10. Pin cathode configuration. (a) Side view photograph of NGD and SOP on the water surface, with a 5 mm gap. (b) Light intensity profile along the z-axis of the SOP, with FWHM representing the SOP thickness. (c) Plasma layer thickness, either CUP or SOP, as a function of the gap distance.*

Similarly, SOP diameter is measured by photographing NGD at 45°, as illustrated in **Figure 11a**. Image processing techniques yield an intensity profile (**Figure 11b**) which is more complex than the one for thickness due to the presence of multiple specks and filaments. The FWHM of the profiles permits to determine the diameter of CUP and SOP. The resulting diameters, plotted against the gap in

**Figure 11c**, expands linearly from 2.2 mm to 4.7 mm, reaching a maximum value of 5.5 mm at a 7 mm gap, after which it decreases to about 2.5 mm at a gap of 12 mm. This indicates a contraction of the plasma layer as it transitions back to a CUP.

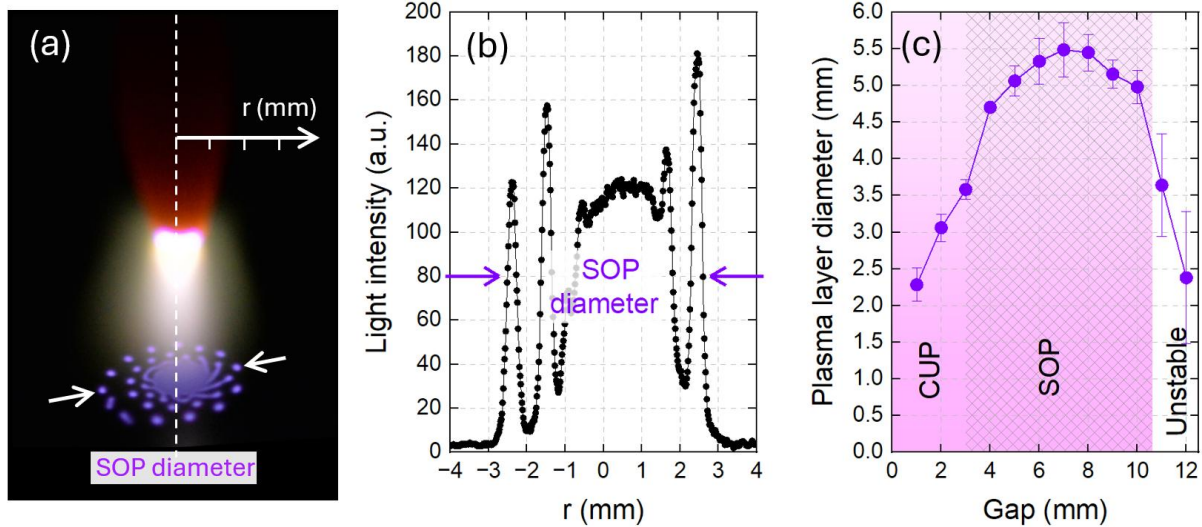

**Figure 11. Pin cathode configuration. (a) Side view photograph of NGD and SOP on the water surface, with a 5 mm gap. (b) Radial light intensity profile of the SOP measured from panel (a), showing a multi-peak distribution, with FWHM representing the SOP diameter. (c) Plasma layer diameter, either CUP or SOP, as a function of the gap distance.**

### 3.2.3. Influence of electric field (gap = 8 mm)

Since SOPs exhibit optical and spectral properties distinct from those of NGD, it is reasonable to hypothesize that their electrical properties also differ. To test this hypothesis, the reduced electric field (E/n) is estimated within the interelectrode gap (i.e., from the pin electrode to the water surface) using the Paris method, which determines E/n from the intensity ratio of the spectral lines of $N_2^+$ and $N_2^+$ at 391.4 nm and 337.1 nm, respectively, measured in dry air [Paris, 2005].

In this study, NGD forms above a heated water surface at temperatures exceeding 370 K (**Figure 6b**), leading to a high discharge environment with RH = 94%. Under these conditions, the excitation of nitrogen molecules to $N_2$ ($C^3\Pi_u$) and $N_2^+$ ($B^2\Sigma_u^+$) is not solely governed by electron impact. Indeed, it is significantly influenced by collisional quenching due to $N_2$, $O_2$ and particularly $H_2O$ molecules, as detailed in Section 2.2.4. Consequently, the Equation (14) expresses $R_{dry}$ (whose value is unknown under our experimental conditions but mandatory to apply the method of Paris) as a function of $R_{wet}$ (the line ratio measured by OES in humid air) and the concentrations of $N_2$, $O_2$ and $H_2O$ (i.e. water vapor molecules) obtained via mass spectrometry. Based on Equation (14), the ratio $R_{dry}/R_{wet}$ is plotted as a function of water vapor concentration in **Figure 12a**. Under 94% relative humidity, a correction factor of $R_{dry}/R_{wet} = 0.64$ is obtained.

As illustrated in **Figure 12b**, the values of $R_{wet}$ vary with axial position $zzz$. By applying the 0.64 correction factor to $R_{wet}$, we obtain $R_{dry}$, which is required for the Paris method. For instance, at z=0, we measure $R_{wet} = 0.53$. Applying the correction factor yields $R_{dry} = 0.34$. By projecting $R_{dry}$ onto the x-axis in **Figure 12c**, we estimate a reduced electric field of 1580 Td at z = 0 mm. This enables the reconstruction of the axial profile of E/**n** along the interelectrode gap, as shown in **Figure 12d**.

The value of 1580 Td at z=0 mm is not directly associated with the plasma but rather attributed to thermo-field electron emission, a synergistic effect of thermionic and field emission processes, which typically occurs in electrode materials at high temperatures [Haase, 2016]. This result is consistent with the estimations of Brisset et al., who report values in the range of 1000–3000 Td for negatively pulsed discharges [Brisset, 2019], and with the simulations of G.V. Naidis, where a value of 10 kTd is considered to account for specific conditions in DC discharges at low pressures and/or in regions near highly curved electrodes [Naidis, 2024]. Beyond this peak, E/n decreases sharply, reaching 480 Td at z = 1. For z = 3–5 mm, it falls below the detection limit before reappearing near the water surface, where it attains a maximum of 425 Td. Similar values have previously been measured for negatively pulsed corona discharges interacting with a metal plate (gap = 18 mm, voltage = 85 kV, pulse duration = 2 ns) [Brisset, 2019].

Using the known particle density, deduced from Equation (2), the axial profile of the absolute electric field is plotted in **Figure 12e**. The field intensity reaches 260 kV/cm at the tip of the pin electrode, decreases to 100 kV/cm at z = 1 mm, before being undetectable in the in the central region of NGD. Then, it increases back to approximately 90 kV/cm near the water surface, where SOPs are formed. To verify whether these values align well with theoretical predictions, the electric field at the electrode tip can be estimated using the point charge model (Equation (19)), where $V_{pin}$ is the applied voltage (−4.5 kV, corresponding to the pin voltage instead of plasma voltage), $r$ is the tip radius of the electrode (0.5 mm), and $k$ is a geometric correction factor, typically ranging between 1 and 3 depending on the electrode's sharpness. Assuming $k = 1$, the estimated field strength is $E = 90$ kV/cm, which closely matches the 100 kV/cm measured 1 mm from the electrode tip (**Figure 12e**). This value also exceeds the breakdown field of ambient air, which is approximately $E_{br} = 30$ kV/cm, confirming that NGD formation is feasible.

$$E = \frac{V}{k \times r} \qquad (19)$$

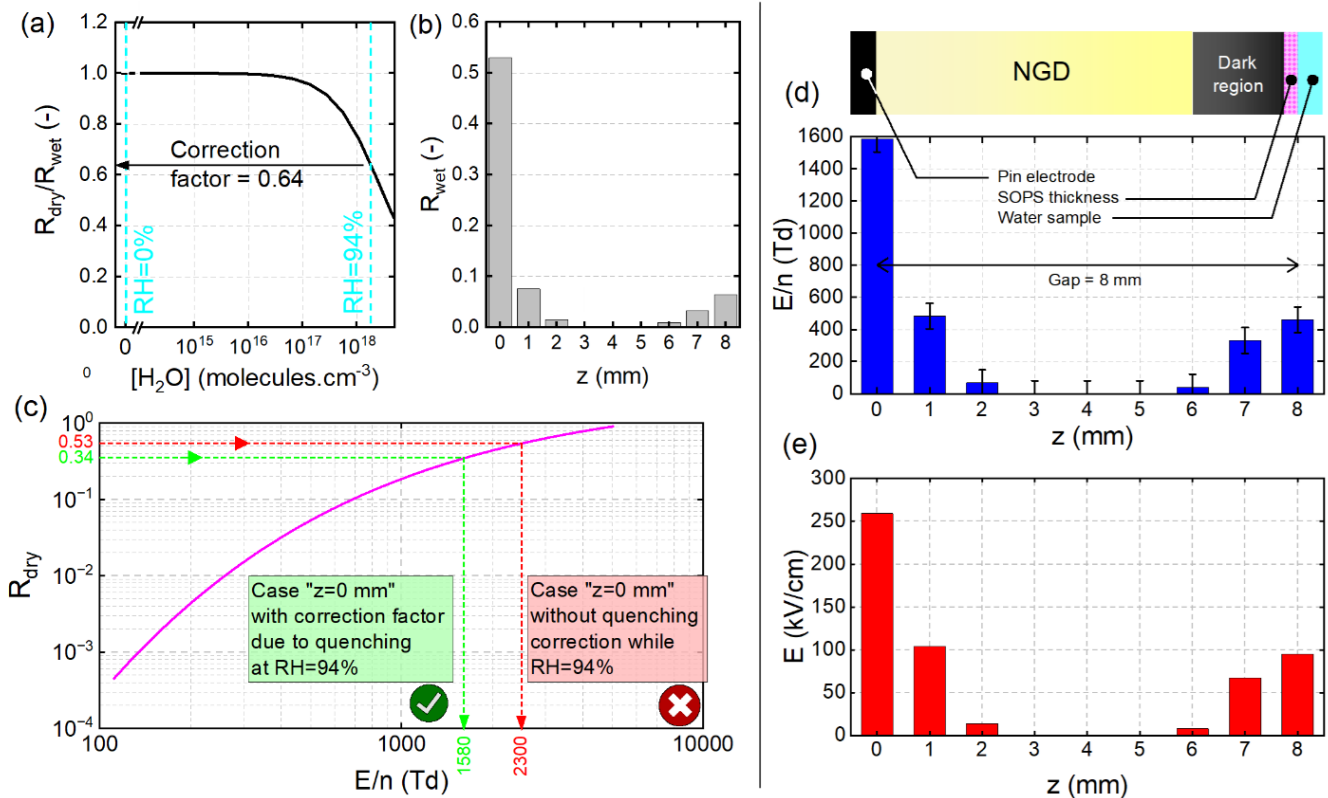

*Figure 12. (a) Influence of water vapor concentration measured by MS in ambient air on the $R_{dry}/R_{wet}$ parameter measured by OES. (b) Axial profile of the nitrogen line ratio ($R_{wet}$) in NGD within ambient air at 94% relative humidity. (c) Nitrogen line ratio in dry ambient air measured by OES as a function of the reduced electric field. This curve is based on [Paris, 2005]. (d) Axial profile of reduced electric field in the 8 mm gap. (e) Axial profile of electric field in the 8 mm gap. $I_d = −30$ mA, $V_d = −1.3$ kV.*

### 3.2.4. Influence of surface tension of water

Similarly to how parameters such as the electrical conductivity of liquids influence the size and complexity of SOP, we aimed to investigate whether surface tension—a less extensively studied parameter—exerts a comparable effect on SOP formation. In this study, the surface tension of tap water, initially measured at 77 mN/m, is systematically reduced by adding varying volumes of an anionic surfactant ($V_{surf}$) to the volume of tap water ($V_{water}$). As shown in **Figure 13a**, a $V_{surf}/V_{water}$ fraction of 1% reduces the surface tension to 42 mN/m, while a 10% fraction lowers it further to 29 mN/m. Correspondingly, the SOP area decreases significantly, with measured values of 17.1 mm², 9.1 mm², and 5.8 mm² for fractions of 0%, 2%, and 10%, respectively.

This reduction in SOP area is accompanied by two notable phenomena: a localized elevation of the liquid surface beneath the pin electrode appearing in **Figures 13e** and **13f** as a water dome and a restructuration of the SOP. At a 3% fraction, the SOP is mostly emissive in the central region, from which highly emissive tips radially emerge. At 10%, this phenomenon is even more pronounced while the tips tend to extend and recover each other.

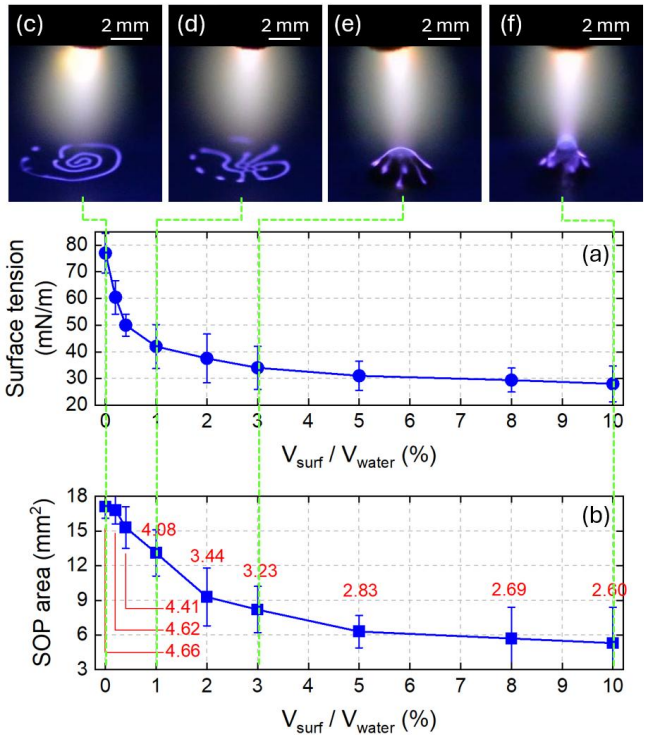

Figure 13. (a) Surface tension of the water sample as a function of the surfactant-to-water volume fraction. (b) SOP area as a function of the surfactant-to-water fraction, with red numbers indicating SOP diameters in mm. (c-f) Photographs of SOP with gap = 8 mm, $I_d = -25$ mA, $V_d = -1.3$ kV. Surface tensions are: (c) 78nmN/m, (d) 42 mN/m, (e) 34 mN/m and (f) 29 mN/m.

# 4. Discussion

The mechanisms underlying the behavior of positive and negative DC glow discharges in interaction with the water surface are analyzed, with particular emphasis on the generation and consumption of active species.

## 4.1. Positive DC glow discharge and CUP: Electron kinetics and OH• radical pathways

As described by Chen et al., PGD can efficiently generate free electrons through secondary ionization of ambient air molecules [Chen, 2002]. These electrons originate primarily from natural sources in the central gap (z = 2-5 mm) as illustrated in **Figure 3a**. Upon creation, these electrons migrate toward the pin anode, where they encounter inelastic collisions with air molecules, resulting in energy loss. In the vicinity of the pin anode, the reduced electric field is so high that it strongly increases electron density. Following Chen et al.'s simulations, a rise from $10^9$ m$^{-3}$ at 300 μm to $10^{12}$ m$^{-3}$ at the electrode surface can hence be obtained [Chen, 2002].

Elevated gas temperatures increase vapor concentration and therefore the availability of H$_2$O molecules from which OH• radicals can be produced in the plasma phase (**Figure 4**). **Table 2** summarizes the main reaction pathways likely to produce these radicals while **Figure 14a** illustrates the spatial repartition of the emissive species within the glow discharge:

- **Near the water surface**: Electrons with kinetic energy exceeding 6.5 eV dissociate H$_2$O into OH• and H• radicals (reaction {1}). Oxygen radicals can also contribute to OH• formation through reaction {2}, although their role is less prominent, as suggested in **Figure 4e**. At the outer edges of the glow discharge (regions with low electric field), electron affinity processes may allow oxygen radicals to capture electrons, forming O⁻ ions (reaction {3a}) which releases 1.46 eV. Then, these O⁻ ions could produce OH• radicals through reaction {3b}. Another mechanism rendering for the production of OH• radicals may combine the production of H$_2$O$_2$ species and the intense heating of the anode pin. More specifically, O$_2$ molecules could form O$_2$•⁻ ions (reaction {4a}), releasing 0.44 eV and subsequently react with water vapor (reaction {4b}) to produce hydrogen peroxide (H$_2$O$_2$) and oxygen. Then, H$_2$O$_2$ can undergo dismutation to form OH• radicals (reaction {4c}) or directly react with O$_2$•⁻ to produce additional OH• radicals (reaction {4d}). reaction {4c} require the presence of catalysts in the liquid phase that could correspond to tungsten microparticles released from the pin electrode subjected to intense heating.

- **In the central region**: The intermediate production of OH• radicals seems to primarily involve electrons via reaction {1}. **Figure 4e** rules out the role of reaction {2}, as no emission from O• radicals is detected. Consequently, reactions {3a} and {3b} do not lead to the formation of OH•. However, reaction {4a} remains possible, allowing O$_2$⁻ to also contribute to OH• radical production.

- **Near the pin electrode**: Although the concentration and energy of electrons are highest near the electrode, water vapor becomes a limiting reactant since it has been consumed upstream. Even if reaction (1) occurs, the scarcity of water molecules limits OH• radical formation, as shown in **Figure 4b**.

Additional reactions, while speculative, are worth noting. For instance, excited N$_2$* molecules could potentially transfer energy to H$_2$O, causing dissociation (reaction {5}). Positive ions like O⁺ could also dissociate H$_2$O (reaction {6}), though the high ionization energy of O⁺ (35.12 eV) makes this unlikely, as no absence of N$_2$⁺ ions in the OES data. Lastly, photodissociation of H$_2$O to produce OH• radicals (reaction {7}) appears improbable due to the required photon energy of 6.5 eV (around 190 nm wavelength), which is not detected in our experiments.

Table 2. Reactions likely to produce OH• radicals in NGD within ambient air.

| Reaction | # | Relevance in our work |
|---|---|---|
| $e + H_2O \rightarrow OH^{\bullet} + H^{\bullet} + e$ | {1} | Yes |
| $O^{\bullet} + H_2O \rightarrow OH^{\bullet} + OH^{\bullet}$ | {2} | Yes |
| $e + O^{\bullet} \rightarrow O^{-}$ | {3a} | Yes |
| $O^{-} + H_2O \rightarrow OH^{\bullet} + OH^{-}$ | {3b} | Yes |
| $e + O_2 \rightarrow O_2^{-}$ | {4a} | Yes |
| $2O_2^- + 2H_2O \rightarrow H_2O_2 + O_2 + 2OH^-$ | {4b} | Yes |
| $H_2O_2 \rightarrow 2OH^\bullet$ | {4c} | Yes |
| $O_2^- + H_2O_2 \rightarrow OH^\bullet + OH^- + O_2$ | {4d} | Yes |
| $N_2^* + H_2O \rightarrow N_2 + OH^\bullet + H^\bullet$ | {5} | Unlikely |
| $O^+ + H_2O \rightarrow OH^\bullet + OH^+$ | {6} | Unlikely |
| $hv + H_2O \rightarrow OH^\bullet + H^\bullet$ | {7} | Unlikely |

The formation of N₂* species at 337.1 nm, can be explained by the first two processes listed in **Table 3**: electron excitation (reaction {7}) and photo-excitation (reaction {8}). The latter involves UV radiation produced during the formation of OH• radicals at 308.9 nm, corresponding to a photon energy of approximately 4 eV. The axial profile of N₂*, shown in **Figures 4a** and **4c**, can be explained by these two reactions:

- **Near the water surface:** N₂* emission arises from both reactions {7} and {8}, driven by energetic electrons and UV radiation, respectively. The proximity to the water surface enhances these processes due to higher local concentrations of electrons and OH• radicals.

- **In the central region of the interelectrode space:** The absence of secondary electrons limits reaction {7}. Additionally, the density of OH• radicals at 308.9 nm is lower compared to the region near the water surface, resulting in weaker UV radiation. Consequently, reaction {8} contributes less significantly here. As a result, N₂* emission in the central region is noticeably lower than near the water surface.

- **Near the pin electrode:** Reaction {8} becomes negligible due to insufficient OH• radical densities and, consequently, minimal UV radiation. However, reaction {7} dominates, supported by higher electron density near the pin electrode (providing ample energy for direct excitation) but also by water vapor which becomes a limiting reactant, hence limiting the production of OH• radicals. As a result, a more significant fraction of electrons can engage in the excitation of N₂*, hence explaining its strong optical emission.

Other reactions that could theoretically produce N₂* are not relevant under our experimental conditions. For instance, reaction {9}, where N₂⁺ ions transfer energy to excite N₂, is ruled out due to the absence of the N₂⁺ band at 391.4 nm in the OES data. Similarly, reaction {10} is unlikely, as the metastable N₂ (A³Σ₀⁺) state, with an energy of 6.17 eV, cannot sufficiently excite the N₂ (C³Πᵤ) and B³Πg states, which require higher energies of 11.03 eV and 7.35 eV, respectively. The absence of Vegard-Kaplan bands (A³Σ₀⁺ → X¹Σg⁺) further invalidates this reaction. Likewise, reaction {11}, involving metastable O₂ (a¹Δg) with a potential energy of only 0.98 eV, lacks sufficient energy to form N₂*. Finally, reaction {12}, which would require secondary electron emission from the water surface, is improbable, as it necessitates bombardment by ions such as O₂⁺ or N₂⁺, or exposure to UV radiation with wavelengths ≤190 nm, which is not observed in our experiments.

*Table 3. Reactions likely to produce N₂* species in ND within ambient air.*

| Reaction | # | Relevance in our work |
|---|---|---|
| $e + N_2 \rightarrow e + N_2^*$ | {7} | Yes |
| $hv_{UV} + N_2 \rightarrow N_2^*$ | {8} | Yes |
| $N_2^+ + N_2 \rightarrow N_2^* + N_2$ | {9} | Unlikely |
| $N_2(A^3\Sigma_u^+) + N_2 \rightarrow N_2^* + N_2$ | {10} | Unlikely |
| $O_2(a^1\Delta_g) + N_2 \rightarrow N_2^* + O_2$ | {11} | Unlikely |
| $hv_{UV} + H_2O \rightarrow OH^\bullet + H^\bullet$ | {12} | Unlikely |

Finally, it is worth mentioning that in pin anode configuration, the absence of N₂⁺ ions, as shown in **Figure 4**, confirms that the electric field is too low to impart electrons with kinetic energies high enough to surpass the first ionization threshold of N₂ (15.6 eV).

## 4.2. Negative DC glow discharge and SOP: space charge dynamics and N₂⁺ ion generation

In the pin cathode configuration, the distribution of species detected by OES differs significantly from the previous case, as illustrated in **Figure 14b**. Notably, the region closest to the pin cathode exhibits the highest light intensity (**Figure 3b**) and the strongest reduced electric field (**Figure 12d**). Indeed, a value as high as 2400 Td is measured at the cathode surface where thermo-field emission enables electron emission by the pin cathode. Then these electrons propagate toward the anode (water surface) while losing energy through inelastic collisions with air molecules.

Some slower electrons accumulate in a subregion of NGD, located 3-6 mm from the cathode, where the reduced electric field drops below the detectable threshold of 100 Td. In this region, while electron densities may be higher than near the cathode, their energies are insufficient to ionize or dissociate air molecules. Instead, these low-energy electrons (depicted as blue circles in **Figure 14b**) attach to oxygen molecules (O₂), forming negative ions such as O⁻ and O₂•⁻ (indicated by # symbols in **Figure 14b**). O⁻ ions have lifetimes of microseconds to milliseconds, while O₂•⁻ ions persist longer, ranging from several milliseconds to hundreds of milliseconds. However, at atmospheric pressure, these lifetimes are too short for the ions to reach the anode. Consequently, negative charges accumulate at approximately 5-6 mm, forming a localized space charge zone. Beyond this zone, in the deeper NGD subregion (6-8 mm from the cathode), optical emission decreases significantly, while E/n rises to 425 Td near the water surface. This intensified field re-accelerates electrons, allowing them to gain sufficient energy to ionize nitrogen (N₂), resulting in the formation of N₂⁺ ions.

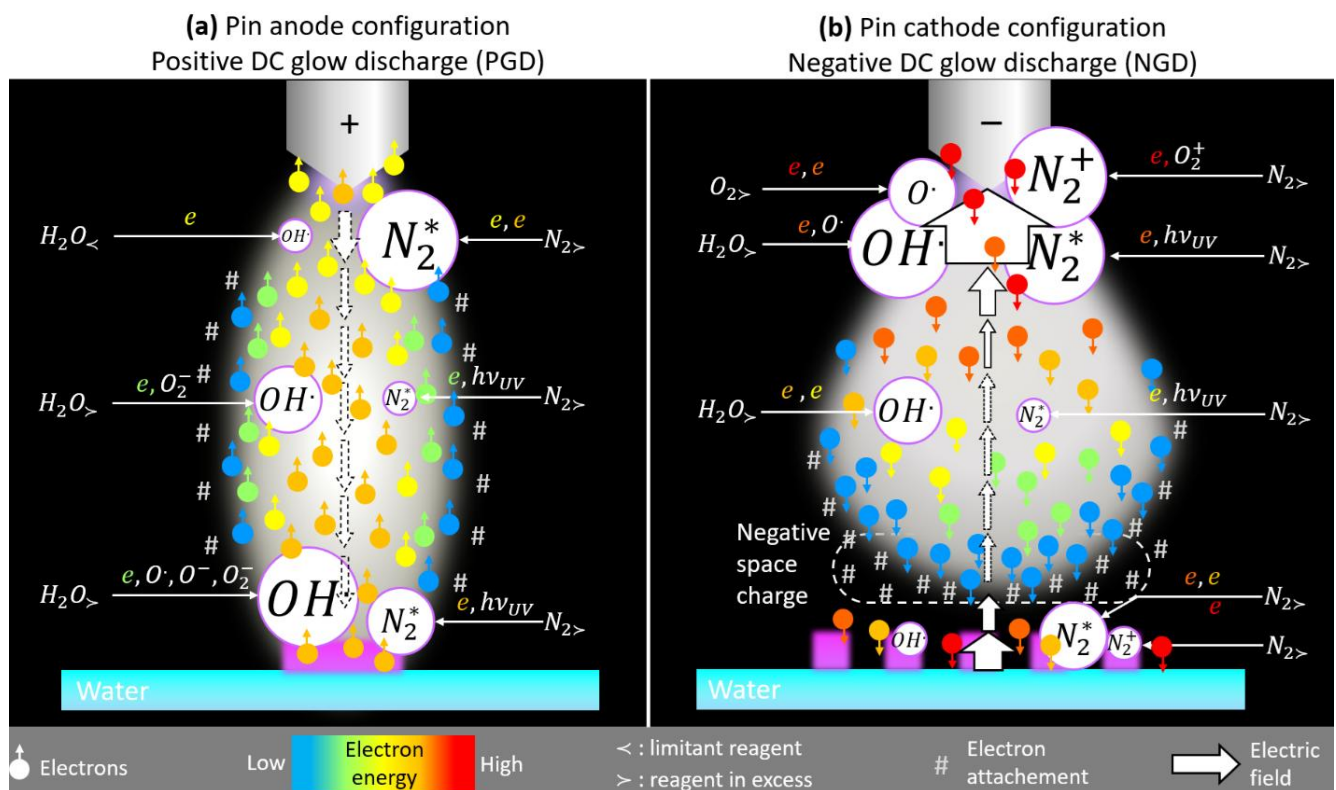

**Figure 14.** *Schematic overview illustrating how a DC glow discharge in ambient air interacts with a grounded water surface, depending on whether the discharge is (a) positive (pin anode configuration) or (b) negative (pin cathode configuration). The proposed mechanisms are based on OES results from Figures 4, 5, 6 and 12.*

The axial profile of OH• radicals, as shown in **Figure 5b**, is radically different from that of the pin anode configuration (**Figure 4b**):

- **Near the water surface:** Reaction {1} still generates OH• radicals, but in much smaller quantities. Despite the abundance of water vapor in this region, the number of electrons with sufficient energy (greater than 6.5 eV) is minimal. Other reactions from **Table 2** are not applicable here, as previously discussed.
- **In the central region of the interelectrode space:** The production of OH• radicals increases in this region, driven by electrons with higher kinetic energies that participate in reaction {1}.
- **Near the pin electrode:** This region exhibits the highest emission of **OH• radicals**, as detected by OES. The electrons here are at their most energetic, enabling significant radical formation. Additionally, the emission of these radicals is accompanied by UV radiation around 308.9 nm (approximately 4 eV). Both reactions {1} and {2} are active in this region, as O• radicals are detected by OES. However, reactions {3} and {4}, which depend on oxygen's electron affinity, are unlikely since low-energy electrons required for these reactions are absent near the pin electrode.

The formation of $N_2$* follows a similar axial profile to that observed in the anode configuration and thus does not require further discussion.

The formation of $N_2^+$ ions is a specific feature of the cathode configuration, closely tied to the presence of a sufficiently intense electric field. Specifically:

- **Near the water surface:** The reduced electric field of 425 Td is strong enough to generate $N_2^+$ ions, as shown in **Figure 5c**. These ions can form via electron impact, either through direct ionization of $N_2$ (reaction {13}), which requires electrons with energies of 15.6 eV, or through a two-step mechanism. In this mechanism, electrons first excite $N_2$ to a higher state (reaction {14a}), followed by ionization of the excited state (reaction {14b}), which requires electrons with approximately 12 eV.
- **In the central region of the interelectrode space:** No $N_2^+$ ions are detected, as the reduced electric field is too weak.
- **Near the pin electrode:** The intense electric field in this region leads to strong $N_2^+$ band emissions, as observed in **Figure 5c**. The high density of $N_2^+$ ions is primarily explained by reaction {13}. Additionally, $O_2^+$ ions formed by electron collisions with kinetic energies around 12 eV (reaction {15}) may undergo charge transfer, enabling the formation of $N_2^+$ ions via reaction {16}. However, photoionization of $N_2$ to produce $N_2^+$ through reaction {17} is unlikely, as no radiation with wavelengths below 90 nm is detected in the experimental conditions.

**Table 4.** Reactions leading to $N_2^+$ ion production in NGD within ambient air.

| Reaction | # | Relevance in our work |
|---|---|---|
| $e + N_2 \rightarrow N_2^+ + e$ | {13} | Yes |
| $e + N_2 \rightarrow N_2^* + e$ | {14a} | Yes |
| $N_2^* + e \rightarrow N_2^+ + 2e$ | {14b} | Yes |
| $e + O_2 \rightarrow O_2^+ + 2e$ | {15} | Yes |
| $O_2^+ + N_2 \rightarrow O_2 + N_2^+$ | {16} | Yes |
| $h\nu + N_2 \rightarrow N_2^+ + e$ | {17} | Unlikely |

## 4.3. Negative DC glow discharge and SOP: influence of surface properties

### 4.3.1. Water surface vs metal surface

The formation of SOPs on metal surfaces has already been demonstrated more than 1 century ago with the pioneering works of G. M. J. Mackay who utilized an anode glow of helium in a discharge tube at a pressure close to 30 mbar [Mackay, 1920]. Then, the physics ruling the self-organization of the patterns has been simulated by Bieniek et al. based on conservation and transport equations of ion and electrons, drift-diffusion and local-field approximations, and the Poisson equation. [Bieniek, 2018]. However, to the best of our knowledge, the formation of SOPs in ambient air on metal surface has not been experimentally observed so far. As already demonstrated by Yan et al., substituting the water surface by a metal surface results in the formation of CUP instead of SOP [Yan, 2015]. Several factors may explain why even in a pin cathode configuration, SOP cannot form on a solid surface at atmospheric pressure.

A first critical difference lies in the thermal dynamics of metal and liquid surfaces. When a metal surface absorbs heat from NGD, it does not trigger evaporation or convective flows. Conversely, a water surface absorbs heat and transfers it deeper into the liquid, resulting in substantial temperature gradients [Brubaker, 2019]. These gradients promote evaporation, as obtained in **Figure 7**, leading to local variations in air density and convective flows. In turn, such variations could induce localized changes in plasma density at the water surface, facilitating the development of self-organization.

Additionally, the chemical processes driven by plasma interactions differ markedly between the two surfaces. On metal surfaces, plasma-induced reactions are relatively simple, typically involving elementary oxidation and corrosion processes [Bigelow, 1988]. In contrast, water surfaces support more complex reactions, generating reactive species such as hydroxyl radicals (OH•), solvated electrons and ions (H⁺, $H_2O^+$ and $NO_2^-$) [Volkov, 2021]. These species can diffuse into the liquid bulk or participate in further reactions. Moreover, plasma ion and electron impacts on the water surface can potentially induce secondary electron emissions, although this effect is unlikely in the current configuration due to the orientation of the electric field toward the pin electrode, which prevents $N_2^+$ ion bombardment.

Finally, the distinct electrical properties of the surfaces could further explain the differences in plasma behavior. Metal surfaces exhibit uniform electrical conductivity, minimizing local charge and electric field variations. Conversely, water, as a polar ionic conductor, is susceptible to polarization under high reduced electric fields (e.g. E/n = 425 Td) [Gruen, 1983] so that its surface could lead to localized variations in conductivity. The presence of surfactants may intensity even more this polarization effect by reducing surface tension, hence altering self-organization of the plasma layer.

### 4.3.2. Water surface vs water-surfactant surface

The experiment investigating the influence of surface tension on SOP characteristics (**Figure 13**) highlight four key effects: the influence of surfactants on tap water, the impact of electric field (without plasma) on water, the impact of electric field on water-surfactant mixture, and finally the effect of D⁻ on the same mixture.

**(i) Effect of the surfactant on tap water:** in pure water, $H_2O$ molecules are randomly oriented, interacting through cohesive forces, such as hydrogen bonds and Van der Waals interactions. When an anionic surfactant is introduced, the amphiphilic surfactant molecules disrupt this arrangement [Jardak, 2016]:
- The hydrophilic heads, negatively-charged, interact strongly with water molecules, forming hydrogen bonds and attracting cations in solution. This can also create a stabilizing counter-ion layer.
- The hydrophobic tails, electrically neutral, are repelled by water and orient themselves toward the air-water interface or cluster together, minimizing exposure to the aqueous environment.

This orientation of surfactant molecules reduces the cohesive forces at the water surface because the hydrophobic tails interfere with the hydrogen bonding between water molecules. As a result, the surface tension decreases, allowing the water to spread more easily, form bubbles, or wet surfaces with less resistance.

**(ii) Effect of the electric field (without plasma) on water:** When a sub-breakdown voltage is applied between the pin electrode and the water surface, an electric field is generated. This field causes water's dipole molecules to reorientate so that the positive charges (hydrogen atoms, δ⁺) of $H_2O$ align towards the electrode, while the negative charges (oxygen atoms, δ⁻) are repelled. As a result, molecular network is restructured at the interface. Electric field is the most intense at the interface, hence locally reducing surface tension, and then decays with depth into the liquid.

**(iii) Effect of the electric field (without plasma) on the water-surfactant mixture:** the electric field amplifies the effects of the surfactant, further lowering the surface tension. The hydrophilic heads of surfactant molecules are strongly attracted to the electric field, while their hydrophobic tails remain directed toward the air. This combined effect causes a localized rise or "dome" in the liquid surface beneath the pin electrode, as observed in **Figure 13**, and significantly alters the liquid's surface properties compared to tap water only.

**(iv) Effect of NGD on the water-surfactant mixture:** If the applied voltage exceeds the breakdown potential, a cold plasma discharge appears. The positively-charged SOP elements, such as specks and filaments, attract the negatively charged hydrophilic heads of the surfactant molecules. This attraction enhances the upward orientation of the hydrophilic heads toward the plasma region, while the hydrophobic tails remain aligned toward the air. This interaction modifies the SOP morphology and further reduces surface tension, resulting in distinctive patterns compared to those formed on pure water surfaces.

# 5. Conclusion

Positive DC glow discharges operate at higher breakdown voltages and power levels, forming circular uniform patterns (CUP) without self-organization. If the pin anode presents a temperature as high as 615 K on its tip, the gas temperature in the interelectrode gap region remains in the 350-370 K range while reduced electric fields is insufficient to generate $N_2^+$ ions and therefore typically lower than 100 Td. OH$^\bullet$ radical production dominates, driven by high electron density near the pin anode and reactions involving $O_2^-$ and $H_2O_2$. $N_2^*$ emissions are influenced by UV excitation and electron impact, peaking near the water surface. PGD thus promotes uniform plasma behavior and chemical activity without structural complexity.

Negative DC glow discharges (NGD) exhibit lower breakdown voltages and enable the formation of SOPs at currents exceeding 15 mA, driven by intense electric fields and specific polarity effects. If the temperature of the pin electrode peaks 650 K, the gas temperature remains in the 350-390 K range in the interelectrode gap. Interestingly, the reduced electric fields peaks at 1580 Td at the pin electrode while a value of 425 Td is obtained at water surface (see **Table 5**). SOP morphology evolves from specks (500 μm) to filaments (6 mm) at gaps of 7–10 mm, with diameters up to 5.5 mm and thicknesses of 210 μm. Surface tension reductions, achieved with surfactants (from 77 mN/m to 29 mN/m), decrease SOP areas from 17.1 mm² to 5.8 mm², inducing localized water domes.

*Table 5. Comparative analysis of CUP and SOP under positive and negative DC glow discharges.*

| | CUP | SOP |
|---|---|---|
| Glow discharge polarity | Positive (PGD) | Negative (NGD) |
| Morphology | Uniform, diffuse spot | Complex, structured (specks, filaments) |
| Typical diameter | 2.2-3.5 mm | 4.5-5.5 mm |
| Thickness | 140-150 μm | 200-220 μm |
| Reactive species in glow discharge | Mostly OH$^\bullet$ radicals A few O$^\bullet$ radicals and $N_2^*$ | Mostly $N_2^*$ and $N_2^+$ ions A few OH$^\bullet$ radicals |
| Electric field strength | < 100 Td | 1580 Td at the cathode 425 Td at water surface |
| Temperature of pin's tip | 615 K | 650 K |
| Plasma gas temperature in the interelectrode gap | 350-370 K | 350-390 K |
| Surface tension | - | Strong influence |

Both PGD and NGD involve plasma dynamics influenced by gas temperature gradients, electric fields, and surface interactions. OH$^\bullet$ radicals and molecular nitrogen emissions are observed in both configurations, although their axial distributions vary. Plasma behavior depends on field strength and energy transfer mechanisms, and water surfaces play a significant role in influencing emissive species and discharge characteristics. In both cases, plasma-liquid interactions contribute to observable chemical activity, albeit with distinct structural outcomes.

The key differences between PGD and NGD lie in electric field strength and resultant patterns, as reported in **Table 5**. PGD displays weaker fields, lacks $N_2^+$ ions, and produces only circular uniform patterns (CUP), while NGD exhibits strong electric fields (1580 Td near the cathode), $N_2^+$ ion generation, and dynamic self-organized patterns (SOP). Additionally, SOP in NGD evolve dynamically, influenced by electrode gap and surface tension, whereas PGD lacks such morphological complexity.

Future research could focus on enhancing SOP stability and control in negative DC glow discharges by exploring novel electrode designs, advanced surfactants, and varying environmental conditions. Additionally, understanding CUP mechanisms in positive glow discharges could optimize applications in uniform plasma treatments, paving the way for broader industrial and environmental uses.

**Author Contributions:** Conceptualization, T.D.; methodology, T.D., E.O.; validation, T.D., E.O.; formal analysis, T.D.; investigation, T.D.; resources, T.D.; writing—original draft preparation, T.D.; writing—review and editing, T.D., E.O.; supervision, T.D. All authors have read and agreed to the published version of the manuscript."
**Funding:** This research received no external funding.
**Data Availability Statement:** The data presented in this study are available on request from the corresponding author due to institutional policies on data sharing.
**Acknowledgments:** The authors acknowledge Sylvain Pledel for the fabrication of the pin electrodes.
**Conflicts of Interest:** The authors declare no conflicts of interest.